\documentclass[twocolumn,showpacs,aps]{revtex4}

\usepackage{graphicx}

\usepackage{amsmath,amssymb}

\begin{document}

\title{Ripples in a string coupled to Glauber spins}
\author{L L Bonilla$^1$, A Carpio$^2$, A Prados$^3$, R R Rosales$^4$}
\affiliation{$^1$G. Mill\'an Institute for Fluid Dynamics, Nanoscience and Industrial Mathematics, Universidad Carlos III de Madrid, 28911 Legan\'es, Spain}
\affiliation{$^2$ Departamento de Matem\'atica Aplicada, Universidad Complutense de Madrid, 28040 Madrid, Spain}
\affiliation{$^3$ F\'{\i}sica Te\'{o}rica, Universidad de Sevilla,
Apartado de Correos 1065, E-41080, Sevilla, Spain}
\affiliation{$^4$Dept. of Mathematics, Massachusetts Inst. of Technology,
77 Massachusetts Avenue, Cambridge, MA 02139, USA}
\date{\today}
\begin{abstract}
Each oscillator in a linear chain (a string) interacts with a local Ising spin in contact with a thermal bath. These spins evolve according to Glauber dynamics. Below a critical temperature, a rippled state in the string is accompanied by a nonzero spin polarization. The system is shown to form ripples in the string which, for slow spin relaxation, vibrates rapidly about quasi-stationary states described as snapshots of a coarse-grained stroboscopic map. For moderate observation times, ripples are observed irrespective of the final thermodynamically stable state (rippled or not).
\end{abstract}


\maketitle
Mechanical systems coupled to spins are used to describe structural phase transitions. Examples include the collective Jahn-Teller effect \cite{kan60}, structural phase transitions with a scalar order parameter exhibiting a central peak in the dynamic response function \cite{sch73} and criticality in martensites and externally driven models \cite{per07}. In many of these models, the mechanical system provides a long range interaction between the spins that produces a phase transition in which the spin polarization ceases to be zero below a critical temperature. While most studies consider the effective spin system obtained after eliminating the mechanical degrees of freedom, it is interesting to focus instead on the effect of the phase transition on the mechanical system. In this paper, we consider mechanical systems coupled to Ising spins that undergo Glauber dynamics \cite{Gl63} in contact with a thermal bath (a single harmonic oscillator connected to Ising spins in the simplest case \cite{PBC10}). There is a phase transition at a critical temperature below which the spin polarization is nonzero and ripples appear in the mechanical system. These thermodynamically stable ripples are inhomogeneous stationary states of the mechanical system, which are quite simple below the critical temperature. On the other hand, there are long-lived dynamical ripples with a wide variety of shapes at any temperature provided the period of mechanical vibrations is short compared to the spin relaxation time. In this limit, the spins are frozen during long time intervals between spin flips and they fix a quasi-stationary state about which the mechanical system oscillates. Observations of the system may consist of time averages over intervals sufficiently long to include many oscillation periods but short compared to the intervals between spin flips. Then these observations will sample a coarse-grained stroboscopic map consisting of successive quasi-stationary states that show ripples. After a much longer time during which sufficiently many spin flips have occurred and due to the dissipation introduced by the Glauber spin dynamics, the ripples eventually evolve to the simple version obtained from the equilibrium thermodynamics of the spin-mechanical system.

These considerations may apply to the evolution of ripples in suspended graphene sheets. Ripples are ondulations of the sheet with characteristic amplitudes and wave lengths that, according to experiments, do not have a preferred direction \cite{mey07}. Time resolved ripples and defects in graphene sheets can be observed using aberration corrected TEMs that collect data every other second, a time much longer than microscopic times such as the one it takes a sound wave to cross one lattice constant \cite{mey08}. As a direct generalization of theories of defect motion in planar graphene \cite{car08}, atom motion in a suspended graphene sheet may be described by the von Karman equations discretized on a hexagonal lattice \cite{BC11}. Coupling the vertical motion of graphene atoms with an Ising spin located at the same lattice point may account for a spontaneous trend of the sheet to bend upwards or downwards. Spin dynamics adds dissipation to the von Karman equations and thus the spin relaxation time should be much longer than microscopic mechanical times. Experimental observations are taken over long time intervals and therefore should correspond to different takes of a coarse-grained stroboscopic map similar to that described in this paper.

{\em Model and rippling phase transition.} Our mechanical system is a chain of oscillators with next-neighbor interaction which becomes a string in the continuum limit:
 \begin{equation}\label{a1}
\mathcal{H}=\sum_{j=0}^N\left[\frac{p_j^2}{2m}+\frac{m\omega^2}{2}(u_{j+1}-u_j)^2- fu_j\sigma_j\right],
\end{equation}
with $u_0=0=u_{N+1}$. The $j$th oscillator is coupled linearly to an Ising spin $\sigma_j=\pm 1$. The spins are in contact with a thermal bath at temperature $T$ and flip stochastically following Glauber's dynamics \cite{Gl63} at temperature $T$. Unlike the case of a regular mass-springs chain, here each triplet is biased against being straight, with $u_{j+1}-u_j = u_j-u_{j-1}\/$. The applied force (whose sign flips at random) makes the ``preferred" state at any instant a wedge shape. This translates the loose idea that the three carbon links each atom shares to build the graphene sheet do not want to be in a plane because of the fourth ``free'' link (which may push the atoms up or down the horizontal planar configuration). The free chemical bonds of the carbon lattice in the graphene sheet may be assimilated to our spins, which interact with phonons modeled by the oscillators. Of course, in order to have a more realistic model of a graphene sheet, the structure of its 2D lattice should be taken into account. Nevertheless, we hope that this simple model will be able to capture the main physical mechanism involved in the rippling of graphene sheets.

\begin{figure}[htbp]
\begin{center}
\includegraphics[width=6cm]{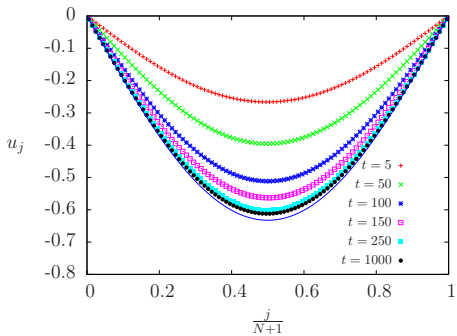}
\caption{Stable ripple state without internal nodes for $\theta=0.9$, $\delta=\alpha/\omega_0=0.1$ and $N=10^4$. For each trajectory, the initial configuration consists of a flat string at rest and random spins, such that the average spin polarization has a sinusoidal shape. Averages over 100 trajectories and spatial averages over 100 oscillators centered at a given one have been performed in order to ensure good averages. See also movie in the Supplementary Material. }
\label{fig1}
\end{center}
\end{figure}

Thus at any time $t$, the system may experience a transition from $(\mathbf{u},\mathbf{p},\mathbf{\sigma})$ to $(\mathbf{u},\mathbf{p},R_j\mathbf{\sigma})$ at a rate given by \cite{Gl63}
\begin{equation}\label{a2}
W_j(\mathbf{\sigma}|\mathbf{u},\mathbf{p})=\frac{\alpha}{2} \left(1-\beta_j\sigma_{j}\right)\! , \quad \beta_j=\tanh\!\left( \frac{f u_j}{k_{B}T}\right)\!,
\end{equation}
where $R_j\mathbf{\sigma}$ is the configuration obtained from $\mathbf{\sigma}$ by flipping the $j$-th spin and $k_{B}$ is the Boltzmann constant. The parameter $\alpha$ gives the characteristic attempt rate for the transitions in the Ising system. Individual spins experience a mutual long-range interaction through their coupling to the string. This long-range interaction causes a phase transition of the spin system: the spins have non-zero polarization for $T<T_c$ whose counterpart for the string is the formation of ripples. To see this, we find the following effective potential by integrating $e^{-\mathcal{H}/(k_B T)}$ over the spin configurations \cite{PBC10}: $\mathcal{V}_{\rm eff}=\sum_{j=0}^N\{m\omega^2(u_{j+1}-u_j)^2/2 -k_B T\ln\cosh[fu_j/(k_B T)]\}$. The extrema of this potential satisfy
\begin{equation}\label{a3}
m\omega^2(u_{j+1}+u_{j-1}-2 u_j)+f\tanh\!\left(\frac{fu_j}{k_BT}\right)\!=0,
\end{equation}
which, linearized about the trivial solution $u_j=0$ (horizontal string) yields
\begin{equation}\label{a4}
U_{j+1}+U_{j-1}-2 U_j + \frac{f^2U_j}{m\omega^2k_{B}T}=0,
\end{equation}
for $j=1,\ldots N$ with boundary conditions $U_0=U_{N+1}=0$. At $T=T_c$, (\ref{a4}) should have a nonzero solution $U_j=e^{ijk}$. Inserting this in (\ref{a4}), we get $f^2/(m\omega^2k_{B}T_c)= 4\sin^2(k/2)$. A linear combination of $e^{ijk}$ and $e^{-ijk}$ that satisfies $U_0=0$ is $U_j=\sin(jk)$. The other boundary condition $U_{N+1}=0$ yields $k_n=n\pi/(N+1)$, $n=1,\ldots, N$. The largest possible critical temperature corresponds to $n=1$ and therefore
\begin{eqnarray}
T_c=\frac{f^2K_N^2}{k_Bm\omega^2},\,\, K_N=\frac{1}{2\sin\!\left(\frac{\pi}{2(N+1)}\right)\!} \sim \frac{N}{\pi} \, (N\to\infty). \label{a5}
\end{eqnarray}
The critical temperature $T_c$ remains finite as $N\to\infty$ if we define a finite $\omega_0$ such that $\omega=\omega_0 K_N$. At $T=T_c=f^2/(m\omega_0^2k_B)$, there is a phase transition so that the trivial solution of (\ref{a3}) is linearly stable for $T>T_c$ and unstable otherwise. We introduce the nondimensional variables $u^*_j=fu_j/(k_BT_c)=m\omega_0^2u_j/f$,  set  $\theta=T/T_c$ and omit the asterisks so as not to clutter the formulas. Below the critical temperature, stable non-uniform states corresponding to ripples in the string appear. The first such state is
\begin{eqnarray}
u_j=\pm 2\sqrt{1-\theta}\,\sin\!\left(\frac{\pi j}{N+1}\right)\!+ O(|1-\theta|),   \label{a6}
\end{eqnarray}
as the reduced temperature $\theta=T/T_c$ tends to 1 from below. This result follows from bifurcation theory for the macroscopic equations for the averages $\langle u_j\rangle$ and $\langle\sigma_j\rangle$ \cite{suppl}. At lower critical temperatures $\theta_n=\sin^2k_1/\sin^2k_n$, $n>1$, other non-uniform states with $n-1$ interior nodes bifurcate from the trivial solution. The non-uniform solutions have a nonzero spin polarization $\langle\sigma_j\rangle\sim\tanh(fu_j/(k_BT_n))$ and therefore the critical temperature $T_c$ is associated with cooperative Jahn-Teller phase transitions, in which coupling to phonons (the string) breaks the symmetry of a doubly degenerated electronic state (the spins) \cite{kan60}. Numerical simulations of the spin-string system confirm this. Below $T_c$, stable string configurations are stationary, nonuniform and exhibit ripples. To test the bifurcation theory, we have performed stochastic simulations at temperatures $\theta>1$ and $\theta=0.9$ for $\delta=0.1$ and $N=10^4$. In Figure \ref{fig1}, we show how an initially flat string at rest evolves to a state close to (\ref{a6}). The initial conditions are random spins, such that the average spin polarization has a sinusoidal shape, and a horizontal, zero-velocity string profile. A qualitatively analogous behavior is found for the majority of initial spin configurations, but for some of them the unstable flat string configuration is stabilized \cite{suppl}. This is a stabilization of the thermodynamically unstable state akin to the one previously found for a single oscillator coupled to Glauber spins \cite{PBC10}. For lower temperatures a similar stationary state without internal nodes is stable whereas the stationary states with $n-1$ internal nodes that bifurcate from the flat string configuration at temperatures $T_n$ are unstable.

\begin{figure}[htbp]
\begin{center}
\includegraphics[width=6cm]{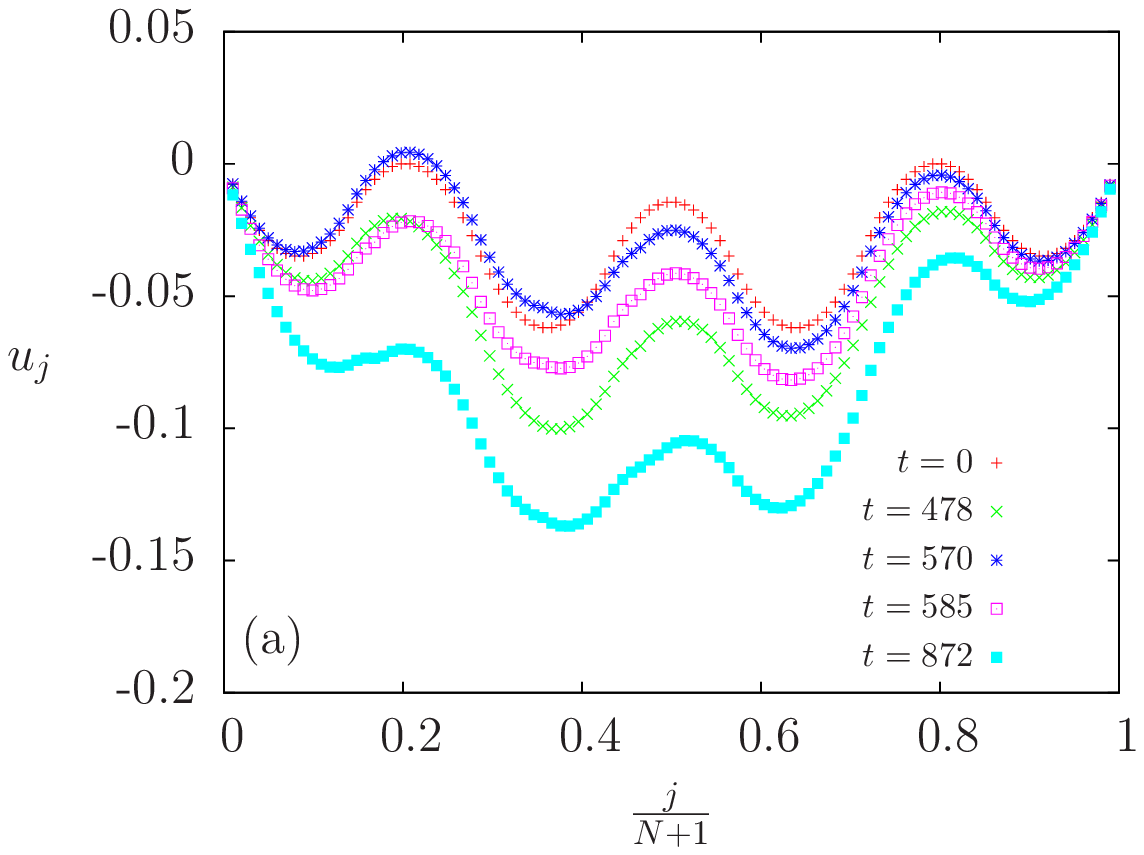}
\includegraphics[width=6cm]{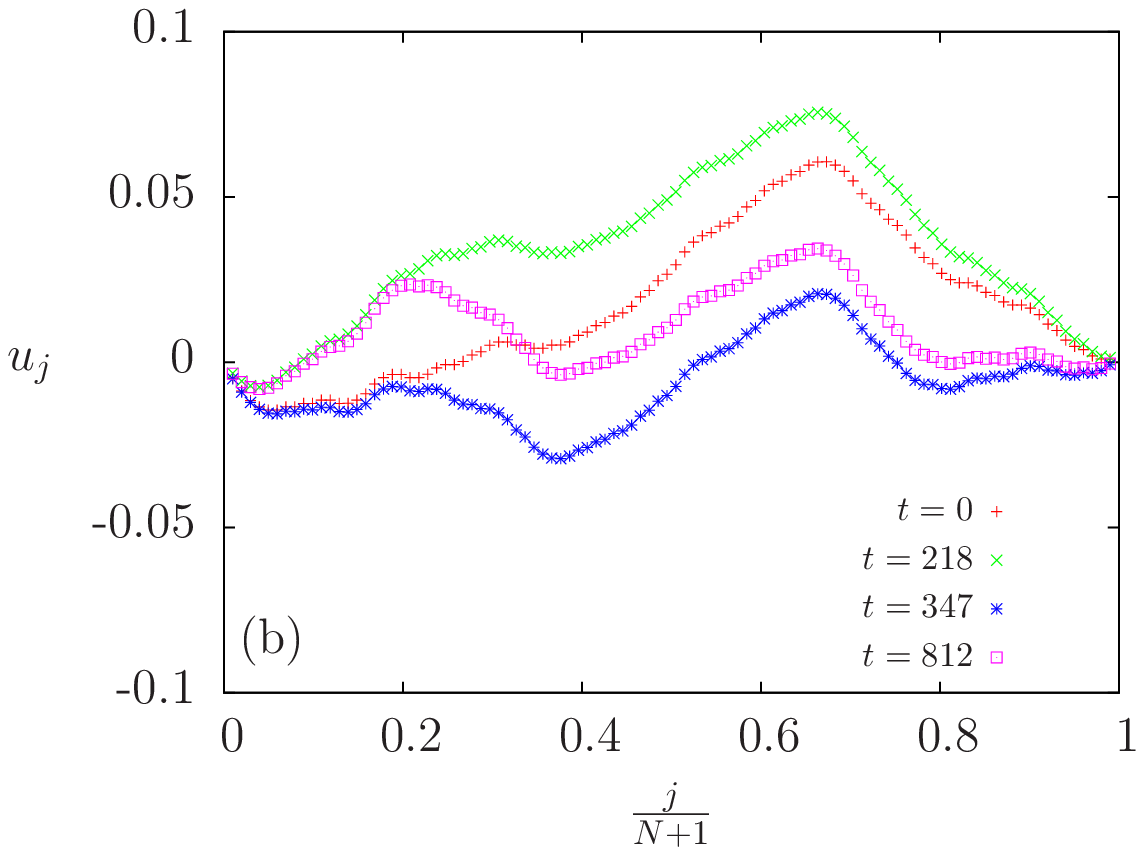}
\caption{Four snapshots of the coarse-grained stroboscopic map for $\theta=0.1$, $\delta= 10^{-4}$ and $N=100$ up to $t=10^3$. The initial configuration consists of a flat string at rest and the spins are (a) initially distributed in seven spin domains of alternating sign or (b) randomly distributed. The times corresponding to the spin flips which change the quasi-stationary string profile are indicated. Ripples with several domains persist for long times whereas the stable configuration corresponding to a string without internal nodes is reached in a much longer time larger than $10^4$. See also movie in the Supplementary Material illustrating how the string oscillates about quasi-stationary configurations given by the coarse-grained stroboscopic map. }
\label{fig2}
\end{center}
\end{figure}

{\em Slow spin relaxation.} Additional insight can be obtained in the limit $\delta=\alpha/\omega_0\ll 1$, in which the spin flip rate is slow compared to the characteristic string frequency. In the following, we will use a dimensionless time $t^*=\omega_0t$, and omit the asterisks as before. Then the $u_j$'s obey the equation of motion
\begin{eqnarray}
\ddot{u}_{j} -K_N^2 ( u_{j+1} + u_{j-1} - 2u_j) = \sigma_j,  \label{a7}
\end{eqnarray}
for $j=1,\ldots,N$ with boundary conditions $u_0=u_{N+1}=0$ ($\ddot{u}=d^2u/dt^2$). The spins $\sigma_j$ are stochastic variables which flip at a rate $W_j(\mathbf{\sigma}|\mathbf{u},\mathbf{p})=\delta(1-\beta_j\sigma_{j})/2$, $\beta_j=\tanh(u_j/\theta)$, $\theta= T/T_c$ instead of (\ref{a2}). Let us consider a trajectory of the system, for given initial states of the string and spins. Since the spin flip rate is very small, the spins are frozen at fixed values during time intervals that are long compared to the longest oscillation period of the string. During the time interval before the spin flip occurs, we may split the solution of (\ref{a7}) in a quasi-stationary and a time-dependent part according to \cite{suppl}:
\begin{eqnarray}
&& u_{j}(t) =u_j^s + v_j(t), \label{a8}\\
&& u_j^s=\frac{1}{K_N^{2}}\!\!\left[ j\!\sum_{l=1}^N\!\left(\!1-\frac{l}{N+1}\!\right)\!\sigma_l-\sum_{l=1}^{j-1}(j-l)\sigma_l\right]\!\!, \label{a9}\\
&&v_j(t)=\sum_{n=1}^N [A_{n}\cos(\Omega_nt) + B_n \sin(\Omega_nt)]\phi_{n,j}, \label{a10}\\
&& \Omega_n=2K_N\sin\!\left(\frac{\pi n}{2(N+1)}\right)\!, \label{a11}\\
&&\phi_{n,j}=\sqrt{\frac{2}{N+1}} \sin\!\left(\frac{\pi nj}{2(N+1)}\right)\!, \label{a12}\\
&& A_n=\sum_{j=1}^N[u_j(0)-u_j^s] \phi_{n,j}, \,\, B_n=\sum_{j=1}^N\dot{u}_j(0)\phi_{n,j}. \label{a13}
\end{eqnarray}
The string profiles (\ref{a8}) represent vibrations of the string about the quasi-stationary configuration (\ref{a9}) whose longest period is $2\pi$ ($\Omega_1=1$ is the lowest frequency). Now let the first spin that flips after $t=0$ be $\sigma_{j_1}$, which changes sign at time $t_1$. Immediately after $t_1$, the right hand side (RHS) of (\ref{a7}) should be replaced by $\sigma_j - 2\delta_{j_1j}\sigma_{j_1}\Theta(t-t_1)$, where $\Theta(x)=1$ for $x\geq 0$, $\Theta(x)=0$ for $x<0$ is the unit step function. The changes in $u_j^s$ and $v_j(t)$ due to the spin flip at $t_1$ are
\begin{eqnarray}
&& \Delta u_j^s\!=\!\frac{2\sigma_{j_1}}{K_N^{2}}\!\!\left[\!\Theta(j-j_1-1) (j-j_1)\!-\!j\!\!\left(\!1-\frac{j_1}{N+1}\!\!\right)\!\!\right]\!\!,\quad\,\, \label{a14}\\
&&\Delta v_j=-\sum_{n=1}^N \!\left(\sum_{l=1}^N\Delta u_l^s\phi_{n,l}\right)\!\phi_{n,j}\cos[\Omega_n(t-t_1)], \label{a15}
\end{eqnarray}
respectively, for $t>t_1$. Successive spin flips produce changes similar to (\ref{a14}) and (\ref{a15}) in the quasi-stationary and time-dependent parts of $u_j(t)$, respectively, at times $t_2,\, t_3,\ldots$ with $t_l-t_{l-1}=O((N\delta)^{-1})$. Time averages over sufficiently long time intervals that are short compared to $(N\delta)^{-1}$ eliminate $v_j(t)$. Thus successive snapshots of averaged string profiles coincide with updated quasi-stationary $u_j^s$ profiles. The latter constitute a coarse-grained stroboscopic map showing how the ripples in the string evolve to their final stable configurations: the horizontal string for $\theta>1$ or a simple parabolic-like profile above or below the horizontal string for $\theta <1$. Figure \ref{fig2} depicts snapshots of the coarse-grained stroboscopic map for an initially flat string at rest with (a) a spin configuration exhibiting seven domains and (b) completely random spins for a temperature $\theta=0.1$, below the critical one. For moderate time intervals stable ripples are observed whereas the stationary configuration of the string without internal nodes is reached at extremely long time intervals. Rippling behavior is also observed for above critical temperatures, $\theta >1$, but the string eventually approaches the flat configuration. In time-resolved experiments such as those with suspended graphene sheets \cite{mey08}, data are taken in long time intervals (1 second), which we use as an estimate for the spin flip attempt rate $\alpha$. Typical times $\omega_0^{-1}$ are 1 ps, therefore $\delta\simeq 10^{-12}$ and the number of spins per linear dimension $N\simeq 10^4$ for square 1 micron samples. Thus,  $1/(N\delta)\approx 10^8$ (much larger than the value considered in Fig.\ \ref{fig2}) and ripple states corresponding to snapshots of the coarse-grained stroboscopic map are observed. The ``true'' thermodynamically stable state would only be reached for extremely long time intervals, much longer than the total observation time in an experiment. On a physical basis, one may expect that the characteristic time associated to the spin flips in graphene be larger than the data-collecting time (1 s), so our estimate for the time between spin flips is actually a lower bound to the actual value. The movie in the Supplementary Material illustrates how the string vibrates rapidly about the quasi-stationary configurations corresponding to successive snapshots of the coarse-grained stroboscopic map. Given the large separation between microscopic times, data collection times and duration of a given experiment, it is important to remark that {\em ripples are observed for all current time-resolved experiments no matter what the temperature and the thermodynamically stable state are. Thus ripples are inherently dynamical and explanations based on thermodynamically stable states do not capture the essence of rippling}.

{\em Continuum limit and fast spin relaxation.} Further analysis of string ripples can be done in the continuum limit $N\to\infty$. From (\ref{a1}) and (\ref{a2}), we obtain equations for averages of $u_j$ and $\sigma_j$. If we split the variables $u_j=\widetilde{u}_j+\Delta u_j$, where $\widetilde{u}_j= \langle u_j\rangle$, set $\widetilde{q}_j=\langle\sigma_j\rangle$ and ignore the fluctuations $\Delta u_j$ in the limit as $N\to\infty$, we get the following nondimensional {\em macroscopic equations}:
\begin{eqnarray}\label{a16}
&&\ddot{\widetilde{u}}_j=K_N^2(\widetilde{u_{j+1}}+\widetilde{u_{j-1}} - 2\widetilde{u_{j}})+ \widetilde{q_j}, \\
&&\dot{\widetilde{q_j}}=  \delta \!\left[\tanh\!\left(\frac{\widetilde{u_j}}{\theta}\right)\!-\widetilde{q_j}\right]\!\! ,  \label{a17}
\end{eqnarray}
for $j=1,\ldots,N$. We now set $\tilde{u}_j(t)=\tilde{u}(x,t)$ with $x=j/K_N$ and take the continuum limit. Then (\ref{a16})-(\ref{a17}) become
\begin{eqnarray}\label{a18}
\frac{\partial^2 \tilde{u}}{\partial t^2}-\frac{\partial^2 \tilde{u}}{\partial x^2}=\tilde{q}, \quad
\frac{\partial \tilde{q}}{\partial t}+\delta \tilde{q}=  \delta\,\tanh\!\left(\frac{\tilde{u}}{\theta}\right)\!,
\end{eqnarray}
to be solved with the boundary conditions $\tilde{u}(0,t)=\tilde{u}(\pi,t)=0$. In the limit $\delta\gg 1$ (fast relaxation of the spins compared to the string time scale), we can approximate the second equation in (\ref{a18}) by $\tilde{q}\approx \tanh(\tilde{u}/\theta) + [\delta\theta \cosh^2(\tilde{u}/\theta)]^{-1} \partial \tilde{u}/\partial t$ and insert this in the first equation. The result is
\begin{eqnarray}\label{a19}
\frac{\partial^2\tilde{u}}{\partial t^2}+\frac{1}{\delta\theta \cosh^2\!\left(\frac{\tilde{u}}{\theta}\right)\!}\frac{\partial\tilde{u}}{\partial t}-\frac{\partial^2\tilde{u}}{\partial x^2}= \tanh\!\left(\frac{\tilde{u}}{\theta}\right)\!,
\end{eqnarray}
whose stationary solutions are (\ref{a6}) just below $\theta=1$. The small damping term in (\ref{a19}) stabilizes the trivial solution above the critical temperature and the stationary ripple solutions below it. The opposite limit of $\delta\ll 1$ has already been studied using the coarse-grained stroboscopic map.

{\em Conclusions.} We have shown that stable ripples appear in a 1D string when each oscillator is coupled to an Ising spin and the latter are in contact with a thermal bath at temperature $T$. Below a critical temperature, the thermodynamically stable string profile is not flat, but nonuniform without internal nodes. In spite of the simplicity of the thermodynamically stable state, more complex ripples appear when the spin flip rate is much smaller than the oscillator period. Although strictly speaking these ripples are evolving in time, they are very long-lived metastable states. The ripples are snapshots of a coarse-grained stroboscopic map depicting the average of the rapid string motion over long time intervals. Whether the final thermodynamically stable is the flat or bent string, ripples should be observed on reasonable time intervals at any temperature.

The system considered here is far from being a realistic model of a graphene sheet. However, the free chemical bonds of the carbon lattice in the latter may be assimilated to our spins, which interact with the phonons modeled by the oscillators. Thus, 2D ripples analogous to the ones found here should appear. This opens the door to understanding the characteristic rippling shown by graphene sheets at any temperature as an  inherently dynamical phenomenon, whose physical mechanism consists of the interaction between free bonds and phonons with widely separated time scales.

This research has been supported by the Spanish Ministerio de Ciencia e Innovaci\'on (MICINN) through Grants FIS2008-04921-C02-01 (LLB), FIS2008-04921-C02-02 (AC), FIS2008-01339 (AP, partially financed by FEDER funds) and FIS2010-22438-E (Spanish National Network Physics of Out-of-Equilibrium Systems), by UCM/BSCH CM 910143 (AC) and by the National Science Foundation grant DMS-0907955 (RRR).

\end{document}